# Nonperturbative Topological Model of Synergistic Phase of 2D Electron Gas Induced by Microwave Excitation

J. C. Phillips

*Dept. of Physics and Astronomy,*

*Rutgers University, Piscataway, N. J., 08854-8019*

## Abstract

The synergistic formation of "zero" (exponentially small or thermally activated) resistance states (ESRS) in high mobility two-dimensional electron systems (2DES) in a static magnetic field B and exposed to strong microwave radiation has attracted great interest. Strong (glassy) disorder apparently stabilizes the new synergistic phase, which can be exactly described by a nonperturbative (nonpolynomial) microscopic topological filamentary model. The model contains many novel features that arise from strong correlations, nanoscopic phase separation, and quasi-one dimensional back scattering.

(jcphillips8@comcast.net)

Two experimental groups have discovered giant longitudinal magnetoresistance oscillations associated with low-field cyclotron resonance in two dimensional electron gases (2DEG) immersed in a microwave bath[1-3]. The new oscillations are qualitatively different from de Haas-Shubnikov (dHS) oscillations produced at high magnetic fields and low temperatures in the absence of the microwave bath (Landau Phase LP), so it is clear that a new synergistic phase (SP) has been produced by the combined interaction of the magnetic and microwave fields. The new SP is observed best in AlGaAs quantum wells with ultrahigh mobilities and carrier densities exceeding $10^{11}/cm^2$. In the SP the resistance minima where $\rho(B) = \rho_m$ virtually erase the residual resistance $\rho = \rho_0$ that arises from background impurity disorder, and are thermally activated with activation energies $\omega_A$ that are as large[3] as 20K, compared to the Landau level spacing $\omega_c$ of 2K.

As discussed below, the evolution of the SP resistance minima as functions of microwave power can best be understood in terms of a first-order phase transition in a microscopically inhomogeneous (glassy) medium.

The synergistic oscillations have attracted widespread theoretical interest (20+ papers[4]), with emphasis on modal instabilities (mainly through microwave-induced mode-mode interactions) of single particle Landau states. It is often argued that these interactions, calculated with mean-field models (or perhaps taking account explicitly of edge states), *might* produce a negative magnetoresistance large enough to cancel the background resistance, leading to zero, or nearly zero, resistance relative to the dc measuring field. While these symmetry-breaking photon-assisted tunneling models are qualitatively correct, they are also incomplete, as they do not explain the origin of the thermal activation energy $\omega_A \gg \omega_c$ and the nature of the SP associated with it. Quite generally, diagrammatic perturbation or mode-coupling models cannot predict first-order phase transitions. In the case of molecular dynamics, the inability of hydrodynamic mode-coupling theories (basically polynomial in nature) to predict ubiquitous stretched exponential relaxation dynamics is well known[5,6]. However, suitable topological models, *not polynomial or perturbative in nature*, successfully predict this relaxation[6], including magic values[7] of the dimensionless stretching fraction β, with an accuracy of a few %. Here we develop a non-polynomial topological model of the synergistic phase that is also highly predictive, and can be used to calibrate sample quality.

Negative magnetoresistance, in itself, while at one time considered to be surprising, is not thought to be so today. Classical negative magnetoresistance (a few %) can occur in the 2DEG because of memory effects and backscattering[8]. A modular (periodic) array of quantum antidots increases the negative magnetoresistance simulated in a classical billiard model by an order of magnitude, to 30+%, and quantum effects measured in the same geometry are about three times larger (90%) than the classical effects[9]. The negative magnetoresistance is associated with modularly enhanced skipping orbits analogous to edge orbits responsible for dissipation in quantum Hall transport in the absence of microwave, where current-induced disorder is observed[10]. The synergistic microwaves are themselves modular, as are the equally spaced Landau levels, so one supposes that the SP represents some kind of mode-locked charge density wave (field-induced order!) that is enhanced by the virtual presence of simultaneously periodic microwave and cyclotron levels. The problem presented to theory is to find ways to describe the mode-locked synergistic state and the activation energy $\omega_A$ without relying on perturbation theory on isolated Landau levels that are separated from the synergistic state by a first order phase transition.

An important feature of the experimental data[11] is that when the oscillations are largest, most symmetric and most numerous, the nodes in $\Delta\rho$ (= $\rho - \rho_0$, where $\rho_0$ is the resistivity in absence of the microwave bath) are fixed and occur at magnetic fields corresponding to level spacings $\omega_c = \omega/(m + j)$, with both $j = 0$ and $j = 1/2$, as sketched in Fig. 1. This implies non-perturbative duality for transitions between Landau bands and impurity bands. This duality arises from electron-hole symmetries that are implicit at the microscopic level, but which can be masked in experiment by space charge effects arising from sample and/or microwave asymmetries that couple to the microwaves because of the absence of inversion symmetry.

First consider the case where the Fermi energy $E_F = (n + 1/2)\omega_c$ is placed at the center of the n th Landau level, $j = 0$, $\omega = m\omega_c$ with $m < 5$. The effect is observed for $n \sim 50$, so terms of order $(m/n)^2$ can be neglected. The area of each Landau orbit is proportional to n, and the radius to $n^{1/2}$. In the normal QHE described by $\rho_0$, the spatial centers of the Landau orbits at the band centers slide in an applied static electric field. The synergistic states that contribute to $\Delta\rho$ are orthogonal to these states and arise from mixing states by the microwave field[12]. There is mixing of n and n + m to create carriers (electrons) or n and n – m to create anticarriers (holes), with the usual dipole symmetry selection rules for interband Landau transitions being broken by the strong electric microwave field.

The transition rates from n to n + m are smooth monotonic functions $f_m(n)$ of $n^{-1}$, so that the difference (here $\delta < \delta_c$)

$$\Delta\rho/\rho_0 = (\rho - \rho_0/\rho_0) = f(n + m) - f(n) + f(n - m) - f(n) = \mathcal{O}(m/n)^2 \qquad (1)$$

Thus $\Delta\rho/\rho_0 \approx 0$ for $j = 0$ transitions between impurity-broadened Landau levels, as derived by several authors[12,13].

Next consider the case $j = 1/2$, where $\omega = (m + 1/2)\omega_c$. The same kind of symmetrical electron-hole pairing that led to cancellation in (1) still holds, even though the states that are involved may no longer be approximated by Landau functions because they involve mixed Landau bands and impurity bands (Fig. 1). Because of this electron-hole symmetry one should find $\Delta\rho(j = 1/2) \approx 0$, but the perturbative calculations[12,13] that gave $\Delta\rho/\rho_0 \approx 0$ for $j = 0$ do not give a similar node at $j = 1/2$. Each of these calculations has assumed that there exists a single-particle density of Landau states that can be convoluted with matrix elements and a symmetry-breaking operator of some kind (a Stark ladder energy and local impurity scattering[12], or an internal local potential step potential barrier[13]). Such convolutions apparently contain hidden inconsistencies when transitions between Landau bands and impurity bands are involved. (One could guess that field-generated extended filamentary impurity band states[14] are omitted in these models.)

In any case, the two sets of periodic nodes at $j = 0$ and $j = 1/2$ are well established in samples where the oscillations are largest, most symmetric and most numerous[11], in other words, in samples where large-scale static inhomogeneities are absent. Even though a function correctly containing both $j = 0$ and $j = 1/2$ sets of nodes cannot be generated by perturbation theory around the $j = 0$ nodes alone, we may confidently assume (in the spirit of Kramers-Kronig transforms) that the correct function satisfies causality conditions that render it analytic. According to the theory of complex variables, the simplest way to represent an analytic function $\Lambda$ having nodes at $z_1, ..., z_m$, is through the polynomial $\Lambda_0 = \Pi(z-z_i)$. If the nodes are evenly spaced, one can also represent $\Lambda$ by a Fourier series $\Lambda_F$, a sum of bounded trigonometric modular functions. The polynomial $\Lambda_0$ is unattractive, because of the high-order singularities at $z = \infty$ (or at $z = 0$ if the function is "normalized" by a factor $z^{-m}$). However, here $\Lambda_F$ is also unattractive, because the observed wave form[11] for $\Delta\rho/\rho_0$ does not resemble the derivative of the density of states $dN(E)/dE$ in Fig. 1, which does resemble a Fourier series in the limit of strong disorder when the Landau bands overlap[15].

The observed wave form $\Delta\rho/\rho_0$ is, however, fitted well by a novel double product analytic function $\Delta\rho/\rho_0 = = As(B)\Lambda_P$, where

$$\Lambda_P = \Pi_m \tanh[(1- \omega/m\omega_c)/T_{Pm}] \, \Pi_m \tanh[(\omega/(m - 1/2)\omega_c -1)/T'_{Pm}] \qquad (2)$$

Here A is a constant, (2) is valid if $\rho > 0$, while if $\rho < 0$, $\Delta\rho/\rho_0$ is replaced by -1; finally, $s(B)$ is an exponential damping factor associated with matrix elements to be discussed below. The product in (2) extends over all integral values of m, both positive and negative, corresponding to quantized energy exchange between the microwave field and the collective current; energy can be either absorbed from the field, or returned to it. The single-particle exclusion principle does not apply to the collective currents of the synergistic phase because the currents are *drift* currents established after multiple one-electron excitations that erase single-particle phase memories. The function $\Lambda_P$ can be regarded as the analytic and topological realization of the configuration space discussion of collective currents in the new phase[1]. The novel double *product* function $\Lambda_P$ cannot be derived by perturbation theory or from polynomial mode-mixing models based on convoluting additive densities of states (or their derivatives). However, it can be understood topologically by analogy with the boundary skipping orbits that are characteristic features of electron motion in strong magnetic fields for ordinary QHE (MICROWAVE power P = 0). There all boundary effects are exponentially localized[16]. In the ultrahigh mobility samples that exhibit synergistic oscillations, for large enough P filamentary paths[14] may be formed providing global channels for drift current flow from electrode to electrode. The electron orbits along these paths will be exponentially localized fragments of Landau orbits resembling braids, as shown in Fig. 2. With this picture one can understand the meaning of $\Lambda_P$. What (2) says is that each m oscillation is topologically correlated to all the other m′ oscillations, in other words, the same drift channels are involved (it is irrelevant whether or not these are edge states). The only difference between m and m′ is a change in size of orbitals involved, while the time-averaged channels, defined by spatial inhomogeneities, including boundaries, contacts, and microwave intensity fluctuations, remain the same, independent of m. Finally, note that in (2), while the nodes are fixed at j = 0 and j = 1/2, the extrema in general do not occur at j =1/4 and 3/4.

The phase of small oscillations with j ~ 1/4 associated with $\rho_{xy}$ resistance minima has previously been explained for small static periodic charge densities in terms of a flat-band condition involving matching of wave function dimensions to charge density periods[17]. Here the similar phases in the giant oscillations of $\rho_{xx}$ can be understood better in the topological context of the channel braiding picture by focusing not on resistance extrema but rather on the nodes of $\Delta\rho/\rho_0$ whose positions are fixed even for very large amplitude oscillations. For (A) j = 0 (1/2) we have circular Landau orbitals of average radius r(n + m (+ 1/2)). The orbitals can be pictured as quantum pearls on a string. If (B) a small part of a larger orbital n + m′ is mixed into a smaller orbital n + m, with m < m′, the elliptical pearls are prolate, are oriented along the string, and overlap[ (j = 1/4). If (C) the converse is the case, that is, a small part of a smaller orbital n + m is mixed into a larger orbital n + m′, the elliptical pearls are oblate, are oriented normal to the string, and are disjoint (j = 3/4). The choice between prolate and oblate is one of sign, and the sign is not easily determined topologically. It is given correctly for j = 1/ 4 by perturbation theory[12] or by analogy with the weakly perturbed static case[17]. Circular skipping orbits at a linear boundary can consist of semicircles, normally incident on the field line. Then (B) is more conductive than (A), because orbital segments span an angle less than π, while (C) is less conductive than (A), as the angles spanned are more than π (Fig. 2). Notice that at j = 0 and j = 1/2 the angle spanned is $\pi$, and the elliptical crescents become semicircles which just contact in both cases. One can also say that the oscillations result from interference between closed and open orbits that occurs when both are present, and is absent at the centers of the Landau and impurity bands, where only one kind of orbit is present.

A characteristic feature of $\Lambda_P$ is that the resistance oscillations are slightly displaced from j = 1/4 (3/4) when the pseudo-Dingle temperatures $T_{Pm} \neq T'_{Pm}$. If $\Lambda_P$ is viewed as a function of a complex variable, then the poles of $\Lambda_P^{-1}$ lie on the real axis and have residues $T_{Pm}$, $T'_{Pm}$. The trends with m of $T_{Pm}$, $T'_{Pm}$ are different[1-3,11], which is expected as the impurity bands and Landau bands have respectively predominantly vector and pseudovector symmetries. The effective order parameters in $\Lambda_P$ are simply the excess filling factors (relative to the band centers) of $\Delta n_m = (1 - \omega/m\omega_c)$ for the Landau bands and $\Delta n_{m+1/2} = (\omega/(m + 1/2)\omega_c - 1)$ for the impurity bands. In principle one can imagine Fourier transforming the spatial areas associated with both kinds of bands; then these filling factors can be identified with the **q** = 0 Fourier component that describes electrode to electrode channels. The exponentials $\exp(\pm\beta\Delta n)$ contained in the hyperbolic tangent functions can be interpreted as drift energies of excess carriers scattered forward or backward by $\mathbf{F}_{microwave}$ along closed or open filaments, respectively, thermally weighted by fictive glassy temperatures $\beta = T_P^{-1}$ that vary depending on n and n + 1/2. The hyperbolic functions in the *product* $\Lambda_P$ cannot be derived by perturbation theory starting from approximations to the densities of states shown in Fig. 1 as trigonometric *sums* (Fourier series).

Forward and backward scattered fragments of open and closed orbits can be paired in various ways to construct extended orbits (Fig. 2 (C). Thus (1) is an open particle, (2) is a closed particle with up orbital momentum, (3) a closed antiparticle with up orbital momentum, and (4) is an open antiparticle. The reader may work out descriptions for (5) – (8) for herself. Because of their vector/pseudovector properties these four fragments are amusingly similar to particle, antiparticle, up and down spin in the Dirac theory of relativistic particles. In principle in special geometries negative resistance is possible in nonequilibrium conditions, but for the samples used in quantum Hall experiments negative resistance is avoided by filamentary formation[18,19].

The matrix elements involved in matching Landau orbital fragments to form open orbits involve the filamentary geometry. The configurational energy gap[1] ~ $T_P$ derived from the microwave drift energy that separates synergistic filamentary open orbits from dHS sliding Landau orbits may be ~10x larger[3] than $\omega_c$, so one should picture the Landau orbital curvature as assisting tunneling across this gap. (The picture of photon-assisted tunneling across Landau gaps[12] associated with internal barriers could be valid in the opposite dHS limit, where $T_P \ll \omega_c$.) Ideally the gap involved in calculating the matrix elements that determine the amplitude of the oscillations in $\Lambda_P$ could be the same as that involved in the period, but this seems unlikely, as even in the Fermi surface oscillations the dHS damping factors do not coincide with the dHvA damping factors[20]. Note that the gap involved in the tunneling matrix elements is more nearly of a local nature, where the fluctuations are larger, than for the gap involved in $\Lambda_P$, which averages over filamentary lengths. The microwave power dependence of the first resistivity minimum exhibits a typical S-shape, reflecting nucleation of drift filamentary channels[1].

The size of the gap involved in constructing extended drift orbits can be estimated from geometrical considerations. Stepping up or down between Landau levels by $\delta E = \omega_c$ reverses the parity of the harmonic oscillator Landau states, corresponding to back scattering (scattering angle $\theta = \pi$). Interpolating between elastic forward scattering, $\delta E = \theta = 0$, and back scattering, a gap $\delta E = \alpha \omega_c$ ($0 < \alpha < 1$) corresponds to scattering by an angle $\theta = \pi/\alpha$. In classical simulations of shortest paths[21], there is a strong tendency towards forward scattering, corresponding[22] to an average $\theta$ much closer to zero than to 1/2 (random walk) or 0.4 (percolating backbone), say $\theta \sim 0.1$, $\alpha \sim 10$. Tunneling probabilities depend slightly on the form assumed for the gap[23], but generally they are given by exp[-($E_g$/F)], where F is an energy derived from the field that is assisting the tunneling to the drift state, here B, with F ~ $\omega_c$ and $E_g$ ~ $\delta E = \alpha\omega_c$ ~ $0.1\omega_c$. We conclude that the tunneling screening factor s(B) in (1) should be given in first approximation by s(B) = exp(-$B_1/\alpha B$), where $B_1$ is the magnetic field associated with $\omega = \omega_c$. It is possible that the damping factor will exhibit small oscillations, here neglected, reflecting the differences between integer and half integer nodes.

In addition to the oscillations in $\rho_{xx}$ discussed here, there are also weak (2% of $\rho_{xx}$) oscillations in $\rho_{xy}$. Those oscillations disappear[1] at filling factors where $\rho_{xx} \to 0$. This behavior is suggestive of a two-phase system containing internal tunneling barriers, with the filamentary phase shorting out the secondary phase when it is highly conductive. Two-phase behavior is commonly observed in strongly disordered glassy systems, and both phases cannot be described simultaneously by perturbation theory. Two-phase behavior will produce macroscopic corrections to Eqn. (2) for $\Delta\rho/\rho_0$. These corrections are not universal and appear to have no microscopic significance; they also appear to be quite small, ~ 2% in conventional sample strips with small contacts, unless they are deliberately enhanced with large contacts or a special microwave configuration, both of which can generate macroscopic filaments and magnetoplasma waves[24]. Fits to the experimental data will be more sensitive to the "synergistic Dingle temperatures" $T_{Pm}$ and $T'_{Pm}$ and to $\alpha$. These should be smooth functions of m, in fact, smoother than the conventional "Dingle temperatures" associated with de Haas-Shubnikov (dHS) oscillations, where there are discrepancies between thermodynamic and transport values[20]. Here the m dependence should be smoothed because the broadening parameters are averaged over optimized interelectrode filaments.

The present topological causality model is not supported by analytic calculations (impossible) or numerical simulations (also impossible without a large quantum computer to match amplitudes and phases in constructing orbits such as those shown in Fig.2). However, in the classical limit (no Landau levels) the relevance of the topological ideas discussed here is apparent in numerical simulations. A strongly disordered granular four-contact model gave large positive magnetoresistance[21] that can be interpreted most easily in terms of a dominant percolative current that follows the filamentary path of least resistance. This macroscopic path corresponds to the first interelectrode path nucleated with increasing microwave power; more paths are formed as power increases above threshold. These paths contain forward and backward segments, and in the limit where the drift energies are large compared to scattering energies, but the system is still in local thermodynamic equilibrium, the macroscopic fractions of the two segments will be given *exactly* by Boltzmann factors. A modular (periodic) array of quantum antidots increases the negative magnetoresistance simulated in a classical billiard model by an order of magnitude, to 30+%, and quantum effects *measured* in the same geometry are even three times *larger* (90%) than the simulated classical effects[25]. [This is unusual, as normally simplified theories predict larger anomalies than are observed experimentally. Here the classical theory omits most of the constructive interference skipping effects of the modular Landau geometry illustrated in Fig. 2.] The negative magnetoresistance is associated with modularly enhanced skipping orbits analogous to edge orbits responsible for dissipation in quantum Hall transport in the absence of microwaves. Finally, studies of shear instabilities in lossless planar hydrodynamic flow have shown that causality imposes limitations on artificial stabilization, which is possible only over a limited range of frequencies[26]. Similarly here, the synergistic phase exists only at low fields, and the Fermi liquid phase is always more stable at high fields.

In conclusion, the novel double product function $\Lambda_P$ provides an excellent description of the main features of the synergistic wave forms observed in high-mobility samples with minimal distortions due to sample and microwave inhomogeneities. Such product forms are characteristic of correlated wave function *amplitudes* (such as are used to describe superfluids and superconductors), but to the author's knowledge have not been applied previously to physical *observables* (that is, *intensities*) such as resistivity. Just as with product wave functions, there appears to be no way to obtain the product $\Lambda_P$ from perturbation theory (photon-assisted tunneling) as a convolution of an additive quasi-particle density of states and its derivative[4,12,27]. These product forms are also suggested by the 1737 Euler identity that expresses the Riemann zeta function in terms of prime numbers[28]. There is a simple topological description, based on duality of closed and open orbits, of the origin of the double product function $\Lambda_P$ that is an extension of the previously discussed filamentary (curvilinear one-dimensional) model of the metal-insulator transition[14]. The strongly correlated product form, dominated by back-scattering, is interpreted as evidence for the persistence of the filamentary spatial geometry[19] with varying magnetic field and band filling (a kind of topologically driven memory effect). Evidence for persistence of localized states from n = 1 to n = 2 Landau states (no microwaves) has been obtained by STM[29]. These robust filaments ($T_P \sim 10\omega_c$) are topological entities that break the symmetry of configuration space without having any special symmetry of their own; one could, if one chooses, call them "strings". They emerge in their ideal form only for high mobility samples of nearly ideally symmetric geometry with few internal tunneling barriers (including point contacts) adiabatically immersed in microwaves. The latter then synthesize the filaments, as sketched in Fig.2. They are the analogue in the synergistic phase of the combined quantization of longitudinal and transverse resistances observed in integer quantum Hall correlations of the Fermi liquid phase[30,31].

I am grateful to R. G. Mani, J. H. Smet, K. von Klitzing, and M.A. Zudov for helpful discussions and access to preprints. R. G. Mani emphasized the need for including damping effects due to configurational tunneling, and he has shown[32] that (2) gives an excellent fit to his data, as well as making valuable comments on the presentation.

**Appendix**

One can easily fit sinusoidal modular functions to data using standard Fourier library routines. The present modular functions, which are not sinusoidal but are hyperbolic because they describe a new non-Fermi liquid phase, require a different approach. There are two remarks that should be useful. First, the first node in eqn. (2) for the oscillating function F occurs not at m = 1, but at m = 1/2. Special care is required in this initial region, where there is a gradual (possibly spinodal) crossover from the dHS Fermi liquid phase to the new synergistic filamentary phase. Second, ideally the nodes occur at integers and half integers. Near these nodes one can derive a simple formula for the curvature of F = Aexp(-bx)Λ near a node at x = m, where m is an integer or half-odd integer, by keeping only three terms in the product Λ, x = m and x = m ± 1/2, and assuming (only a first approximation, to be refined later) that $T_{Pm}$ is the same linear function of m for both integers or half-odd integers. This formula is

$$\delta x F'/F = (1 + 2g\delta x)/ T_{Pm} – b\delta x = 1/T_{Pm} + (2g/T_{Pm} - b)\delta x \qquad (3)$$

where $\delta x = x - x_m$, and $g = \coth(1/2T_{Pm})/\tanh(1/2T_{Pm})$. Note the partial cancellation between the carrier density term g and the matrix element term b in the coefficient of the curvature term δx.

**Figure Captions**

Fig. 1. Sketch of the alternating densities of states of Landau bands and impurity bands. For small n the bands may not overlap, but for the large values of n in the synergistic experiments the overlap is large. Possible transitions that leave the resistivity $\rho$ unchanged at its value $\rho_0$ in the absence of microwaves are indicated.

Fig. 2. At the nodal points (A), where $\rho - \rho_0 = 0$, the orbits can be chosen to be circular (symmetric gauge), but the microwave field organizes the orbits as strings of prolate or oblate ellipses near resistivity minima (B) or maxima (C). In (B) the overlap of the prolate ellipsoids at an angle of less than $\pi/4$ facilitates the formation of field-induced skipping orbits (heavy line). In (C) tunneling (dotted lines) occurs between oblate ellipsoids, or (if the oblate ellipsoids overlap), it is more difficult to form skipping orbits because at the point of overlap the orbital velocities make angles with the instantaneous electric field larger than $\pi/4$.

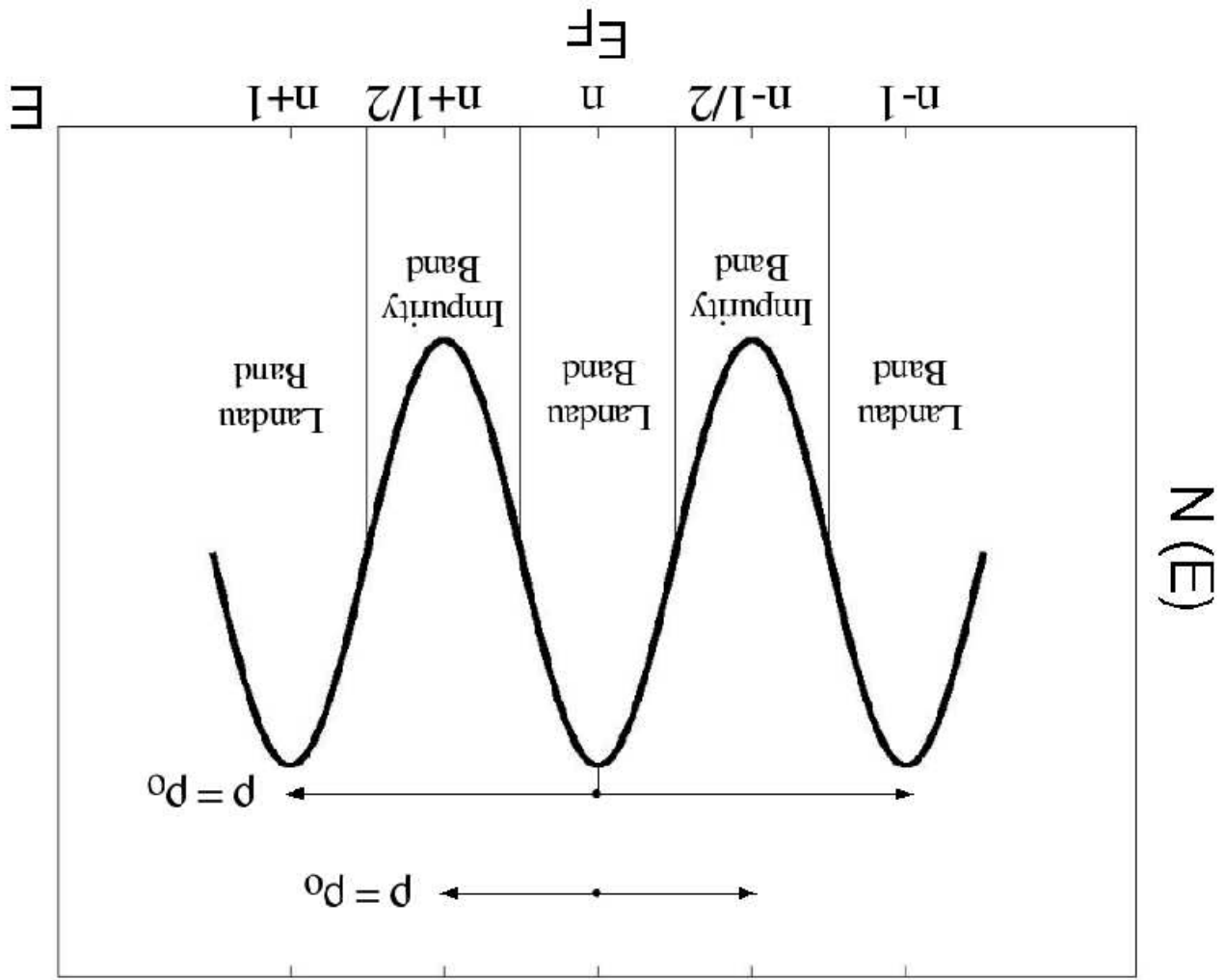

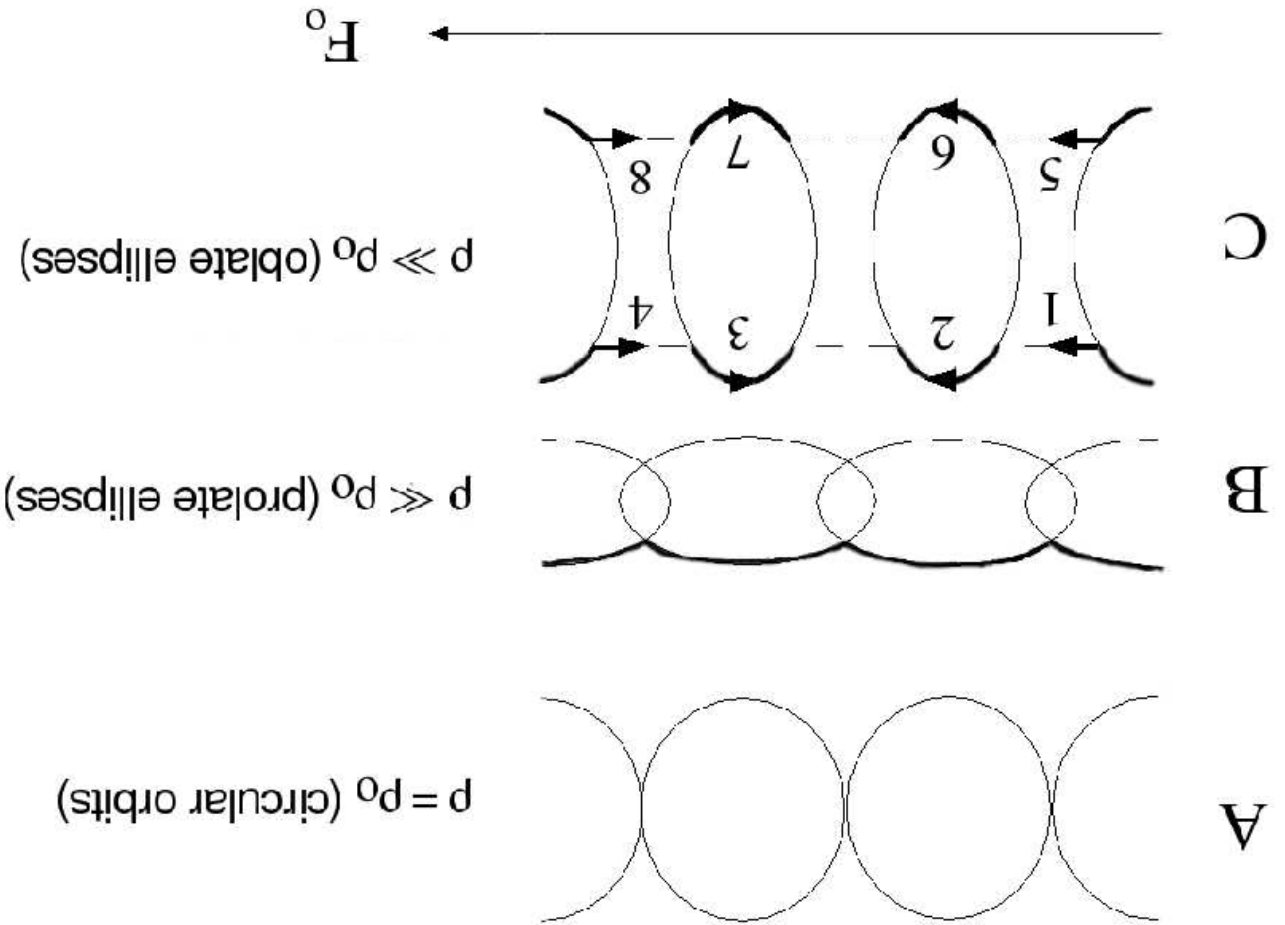

A  $\rho = \rho_o$ (circular orbits)

B  $\rho \ll \rho_o$ (prolate ellipses)

C  $\rho \gg \rho_o$ (oblate ellipses)

$F_o$